\documentclass[twocolumn,showpacs,preprintnumbers,amsmath,amssymb,prl]{revtex4}

\usepackage{graphicx}
\usepackage{bm}

\begin{document}

\title{Van-der-Waals supercritical fluid: Exact formulas for special lines}

\author{V.V. Brazhkin}
\affiliation{Institute for High Pressure Physics, Russian Academy
of Sciences, Troitsk 142190, Moscow Region, Russia and Moscow
Institute of Physics and Technology (State University),
Dolgoprudny 141700, Moscow region, Russia}

\author{V. N. Ryzhov}
\affiliation{Institute for High Pressure Physics, Russian Academy
of Sciences, Troitsk 142190, Moscow Region, Russia and Moscow
Institute of Physics and Technology (State University),
Dolgoprudny 141700, Moscow region, Russia}

\date{\today}

\begin{abstract}
In the framework of the van-der-Waals model, analytical
expressions for the locus of extrema (ridges) for heat capacity,
thermal expansion coefficient, compressibility, density
fluctuation, and sound velocity in the supercritical region have
been obtained. It was found that the ridges for different
thermodynamic values virtually merge into single Widom line only
at $T<1.07 T_c, P<1.25P_c$ and become smeared at $T<2T_c, P<5P_c$,
where $T_c$ and $P_c$ are the critical temperature and pressure.
The behavior of the Batschinski lines and the pseudo-Gruneisen
parameter $\gamma$ of a van-der-Waals fluid were analyzed. In the
critical point, the van-der-Waals fluid has $\gamma=8/3$,
corresponding to a soft sphere particle system with exponent
$n=14$.
\end{abstract}


\maketitle %

A liquid-gas phase equilibrium curve onto the $T,P$ - plane ends
at the critical point. At pressures and temperatures above
critical ones ($P>P_c, T>T_c$), the properties of a substance in
the isotherms and isobars vary continuously, and it is commonly
said that the substance is in its supercritical fluid state, where
there is no difference between a liquid and a gas. An anomalous
behavior of the majority of characteristics are observed in the
vicinity of the critical point. The correlation length for
thermodynamic fluctuations diverges at the critical point
\cite{[1]}; one can also observe a critical behavior of the
compressibility coefficient $\beta_T$, thermal expansion
coefficient $\alpha_P$, and heat capacity $C_p$: the given
properties pass through their maxima under a change of pressure or
temperature. Near the critical point, the positions of the maxima
of these values in the $T,P$ - plane are close to each other
\cite{[1]}. The same is true for the value of density
fluctuations, the speed of sound, thermal conductivity, etc. Thus,
in the supercritical region, there is a whole set of the lines of
extrema of various thermodynamic values. Each of these lines can
be regarded as a "thermodynamical" continuation of the liquid-gas
phase equilibrium curve into the supercritical region. The
smearing and decreasing (in magnitude) extrema of each of the
values form a "ridge" \cite{[2],[3],[4]}. A knowledge of the
positions of the above "ridges" in the $T,P$ - plane is very
important; in particular, it determines a maximum value for such
technologically essential characteristics as the dissolving
ability of a supercritical fluid, the rate of chemical reactions
in a fluid, and others (\cite{[2],[3],[4]} and refs therein). It
turned out that the experimentally observed lines of the "ridges"
are close to an isochores with a slight decrease in density with
increasing temperature \cite{[2],[3],[4]}. Most studies on the
supercritical region focused on examining the "ridge" for the
density fluctuations \cite{[2]}.

G. Stanley suggested the name "Widom line" for the line of the
maximum of the correlation length isotherms and isobars
\cite{[5]}. Since the lines of the maxima near the critical point
merge into one line, the above term was proposed to be used in a
wider sense - in reference to the lines of the maxima of all
values determined by the second derivatives of the Gibbs
thermodynamic potential. The Widom lines for liquid - liquid and
more rarely for liquid - gas transitions for a number of systems
were considered \cite{[5],[6],[7],[8n],[9n]}. It is a priori
unclear how far from the critical point we may say of a single
Widom line for all ridges, and how far from the critical point the
extrema of particular physical quantities can still be followed.
The first part of question may be rephrased as "how far from the
critical point the prefactor that appears in the expression of
fluctuations, and response functions, in terms correlation length
is close to a constant".

Apart from the Widom line, other "special" lines, separating a
fluid state, have been suggested. The Batschinski line \cite{[8]},
sometimes referred to as the Zeno line \cite{[9],[10]},
corresponds to a formal coincidence between the equation of state
for a fluid and the equation of state for an ideal gas.

For solids, an important thermodynamic parameter is the Gruneisen
parameter which reflects the variation of the lattice dynamics
under a change of density. For a fluid, one can introduce a
pseudo-Gruneisen parameter \cite{[12]}, relating such
thermodynamic quantities as the heat capacity $C_v$, thermal
expansion coefficient $\alpha_P$ and compressibility coefficient
$\beta_T$:
\begin{equation}
 \gamma =  \alpha_P / (C_v  \beta_T) . \label{g}
\end{equation}

It is of great interest to analyze the behavior of "special" lines
in a fluid and the pseudo-Gruneisen parameter in the framework of
a simple, exactly solvable model. In the present study we have
analyzed the properties of a van-der-Waals fluid. The
van-der-Waals equation is one of the simplest equations of state
for a fluid. In the reduced variables $T_r=T/T_c, P_r=P/P_c,
\rho_r=\rho/\rho_c$  the equation has the form:
\begin{equation}
(P_r+3\rho_r^2)(3-\rho_r)=8T_r\rho_r. \label{vdw}
\end{equation}

\begin{figure}
\includegraphics[width=8cm]{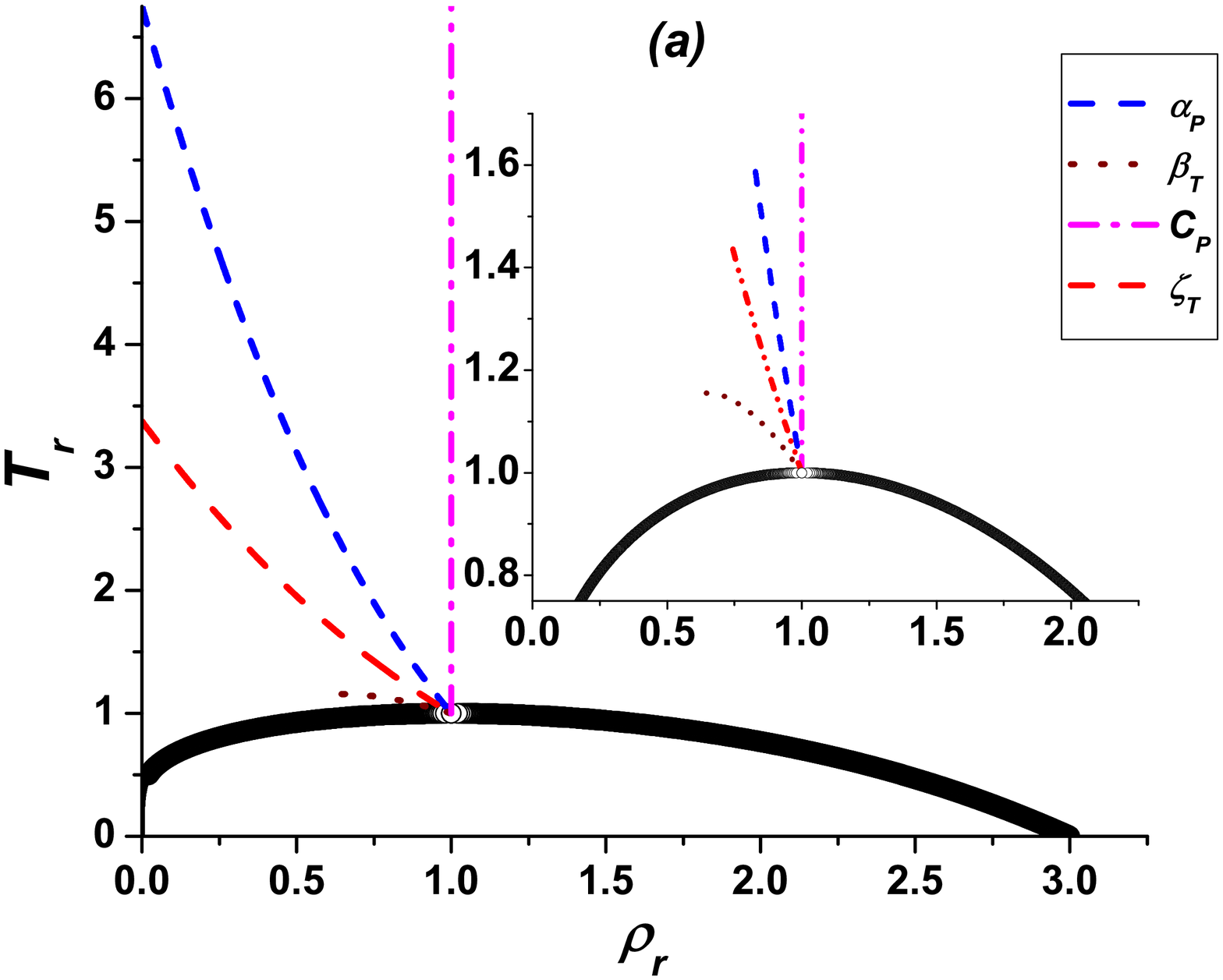}%

\includegraphics[width=8cm]{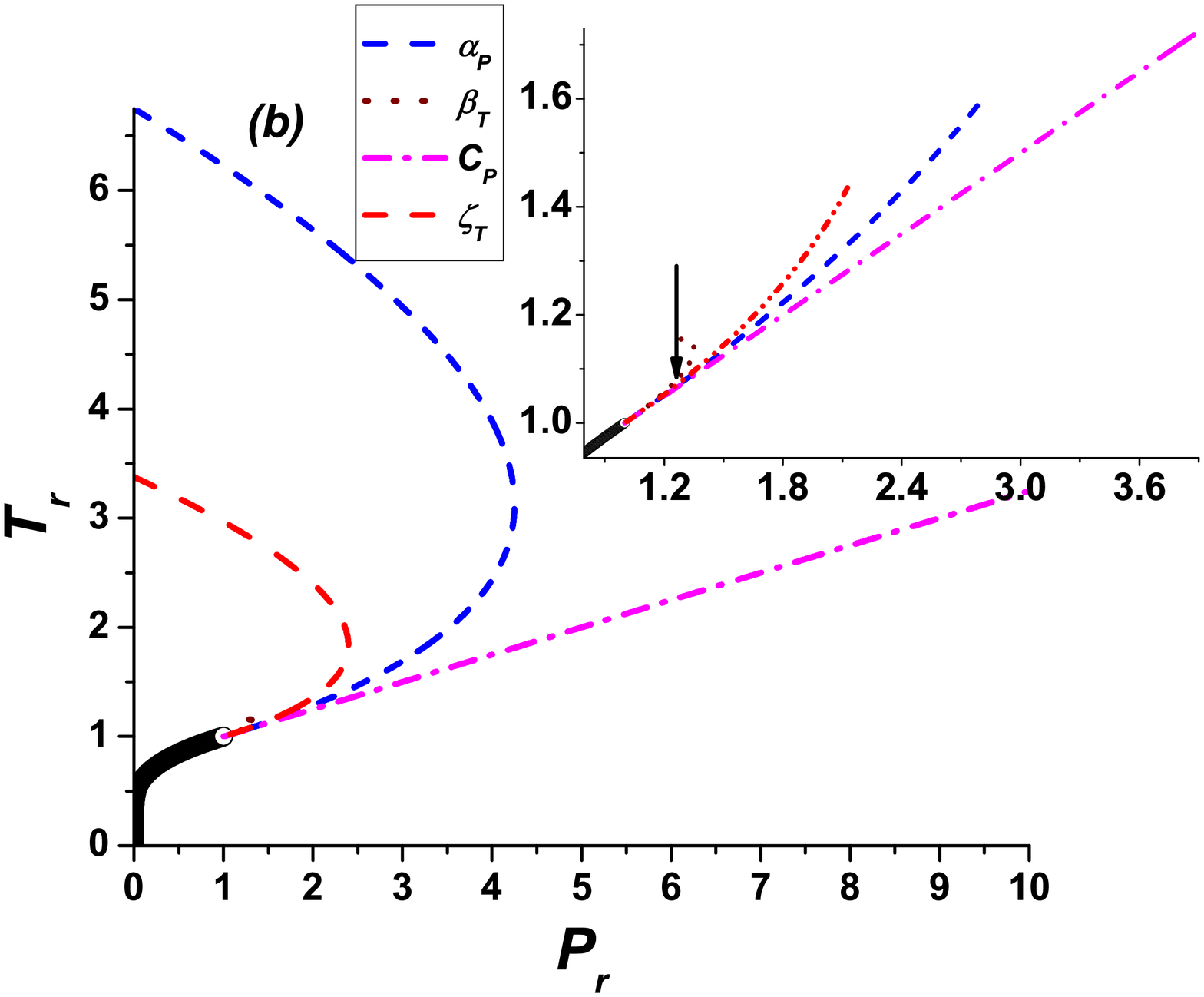}%

\caption{\label{fig:fig1} (Color online) Positions of the maxima
of thermal expansion coefficient  $\alpha_P$, compressibility
$\beta_T$, heat capacity $C_P$ and density fluctuations  $\zeta_T$
in the ($\rho-T$) (a) and ($P-T$) (b) coordinates. The arrow
indicates the approximate end of the "single" Widom line. Thick
lines correspond to liquid-gas transition.}
\end{figure}

It is well known that the van-der-Waals equation does not
reproduces the behavior of real fluids completely and precisely
(\cite{[2],[14]} and refs therein). At the same time the
van-der-Waals equation can be used to understand the fundamentals
of fluid behavior. Besides, this equation is advantageous because
exact analytical expressions can be obtained for most of the
physical quantities \cite{[14]}. At the same time, however strange
it may look, a supercritical region of the $T,P$ - parameters for
the van-der-Waals fluid model has been studied insufficiently. We
know of only one study \cite{[2]} analyzing the behavior of the
line of the maximum of density fluctuations $<(\Delta
N)^2>/<N>=T(\partial \rho/\partial P)_T = \zeta_T$. It is found
that the position of the line of the maxima of density
fluctuations on the isotherms satisfies the equation:
\begin{equation}
\rho_r=3-2T_r^{1/3}. \label{ximax}
\end{equation}
Thus, the above line formally ends at zero density and zero
pressure at $T_r=27/8$ (see Fig.~\ref{fig:fig1}). The speed of
sound has the form:
\begin{equation}
V_s=\sqrt{T/ \zeta_T}=
\left(\frac{6(\rho_r(3-\rho_r)^2-4T_r)}{\rho_r(3-\rho_r)^2}\right)^{1/2}.
\nonumber
\end{equation}
The line of the minimum of the speed of sound on the isotherms
obviously corresponds to the equation (\ref{ximax}). In \cite{[2]}
it was supposed that the extrema of other thermodynamic quantities
in the isotherms would lie roughly on the same line. However, as
we will show below, all "ridges" diverge as one goes even slightly
away from the critical point.

Isothermal compressibility in the framework of the van-der-Waals
model has the form:
\begin{equation}
\beta_T=\frac{1}{\rho}\left(\frac{\partial \rho}{\partial
P}\right)_T=-\frac{(\rho_r-3)^2}{6\rho_r(-4T_r+\rho_r(\rho_r-3)^2)}.\nonumber
\end{equation}
The line of the maxima of compressibility $\beta_T$ on the
isotherms satisfies the equation:
\begin{equation}
T_r=\frac{\rho_r(3-\rho_r)^3}{2(3+\rho_r)}.  \label{(3)}
\end{equation}
This line ends at its own critical point at $T_r = 1.156, P_r
=1.285, \rho_r = 0.646$ (see Fig.~\ref{fig:fig1}). Using a known
thermodynamic relation \cite{[1]}, we obtain:
\begin{equation}
C_P-C_V=\frac{32T_r}{3(4T_r-9\rho_r+6\rho_r^2-\rho_r^3)}.\nonumber
\end{equation}
We remind that at $T_r>1$ $C_V=3/2$. It can easily be seen that
the line of the maxima of the heat capacity $C_P$ in the isotherms
coincides with the isochore $\rho_r = 1$ and is described in the
$T,P$ coordinates by the equation:
\begin{equation}
T_r=\frac{3}{4}+\frac{1}{4} P_r, \label{(4)}
\end{equation}
i.e., represents a direct continuation of the gas-liquid
equilibrium line \cite{[14]} (see Fig.~\ref{fig:fig1}). The
thermal expansion coefficient has the form:
\begin{equation}
\alpha_P=-\frac{1}{\rho}\left(\frac{\partial \rho}{\partial
T}\right)_P=\frac{4(\rho_r-3)}{3\rho_r(\rho_r-3)^2-12T_r}.
\nonumber
\end{equation}
The line of the maxima of the thermal expansion coefficient
$\alpha_P$ in the isotherms corresponds to the equation
\begin{equation}
T_r=(3-2\rho)(3-\rho)^2/4. \label{(5)}
\end{equation}
This above line ends at zero pressure and zero density at
$T_r=27/4$ (see Fig.~\ref{fig:fig1}).

\begin{figure}
\includegraphics[width=4.5cm]{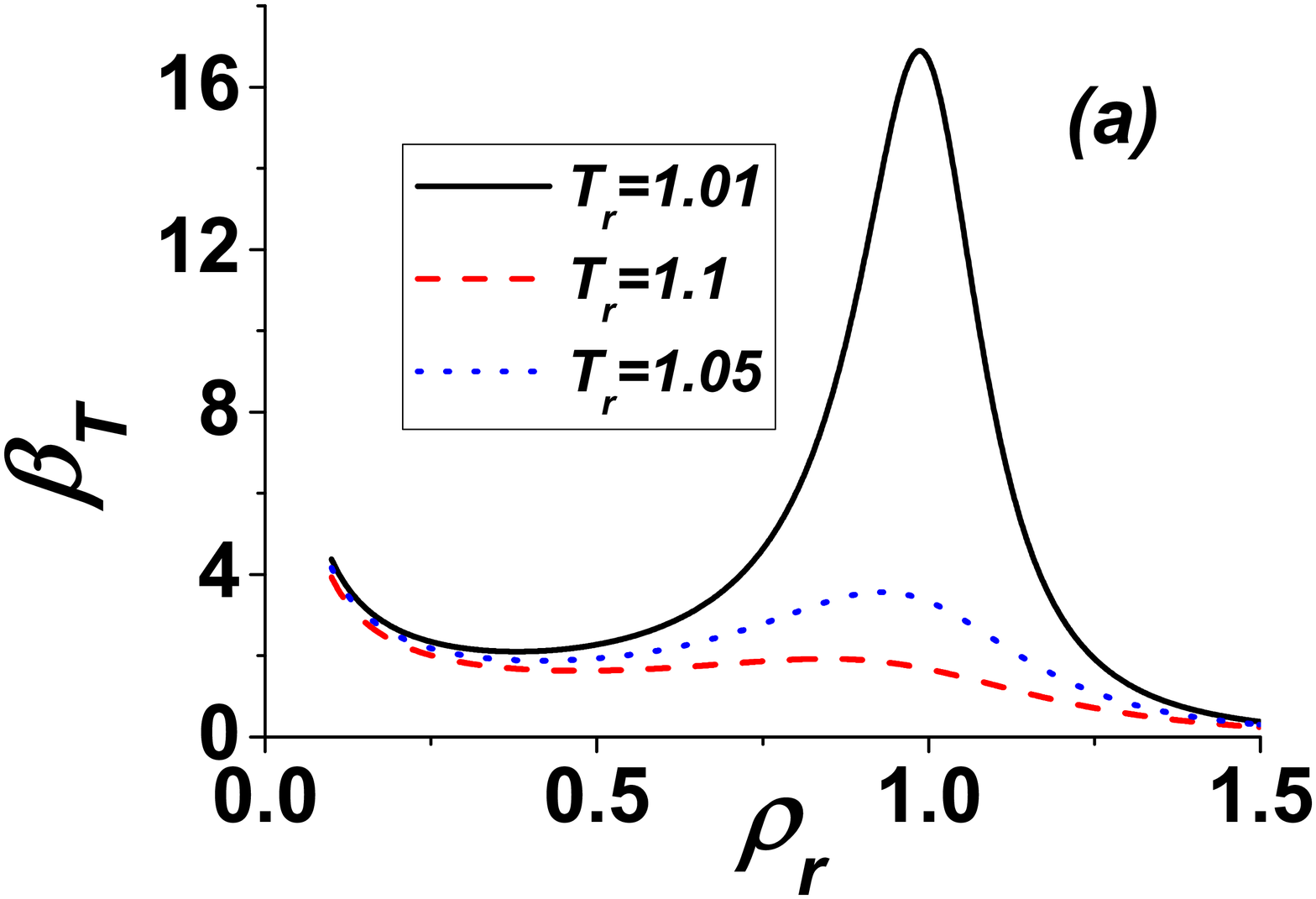}%
\includegraphics[width=4.5cm]{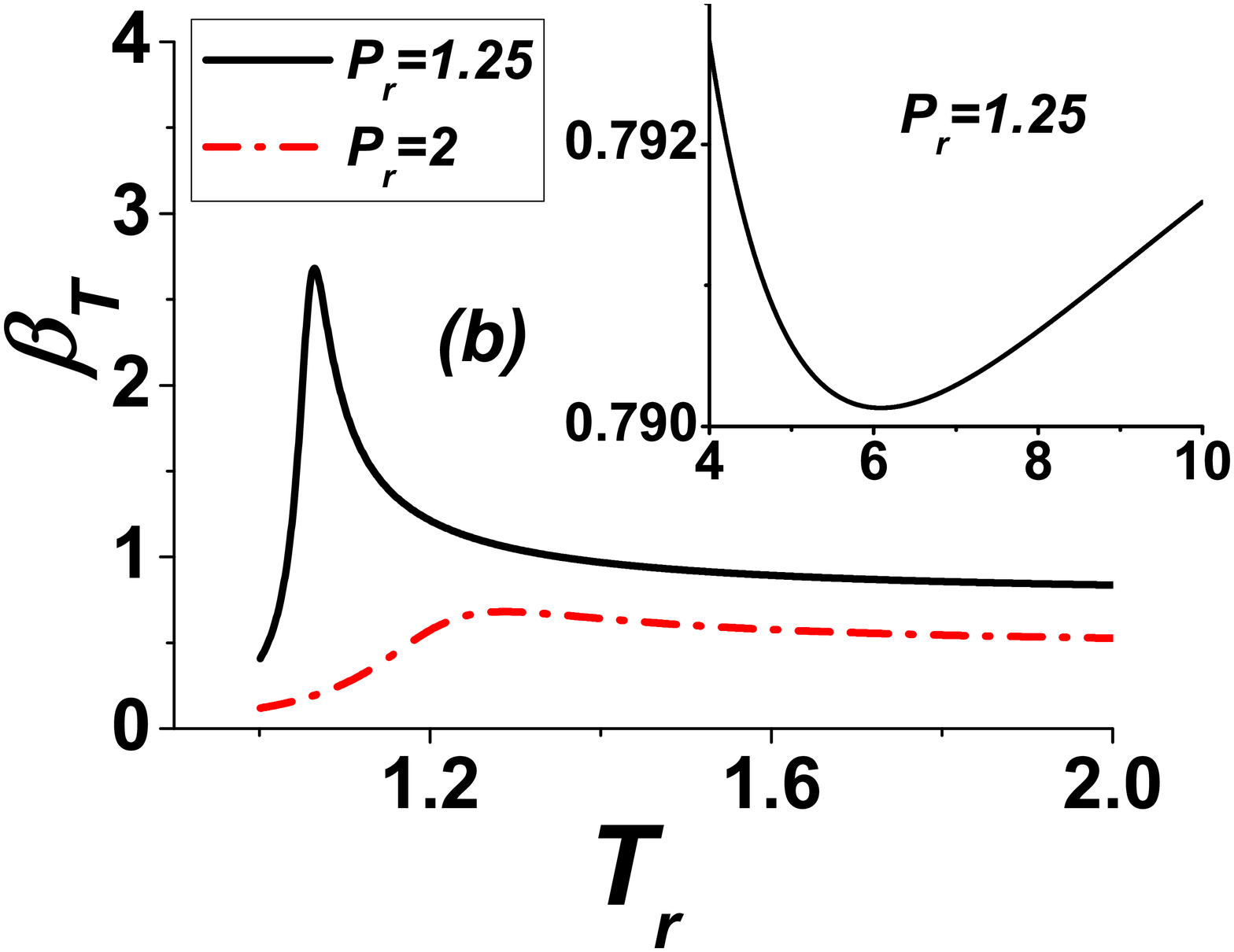}%

\includegraphics[width=4.5cm]{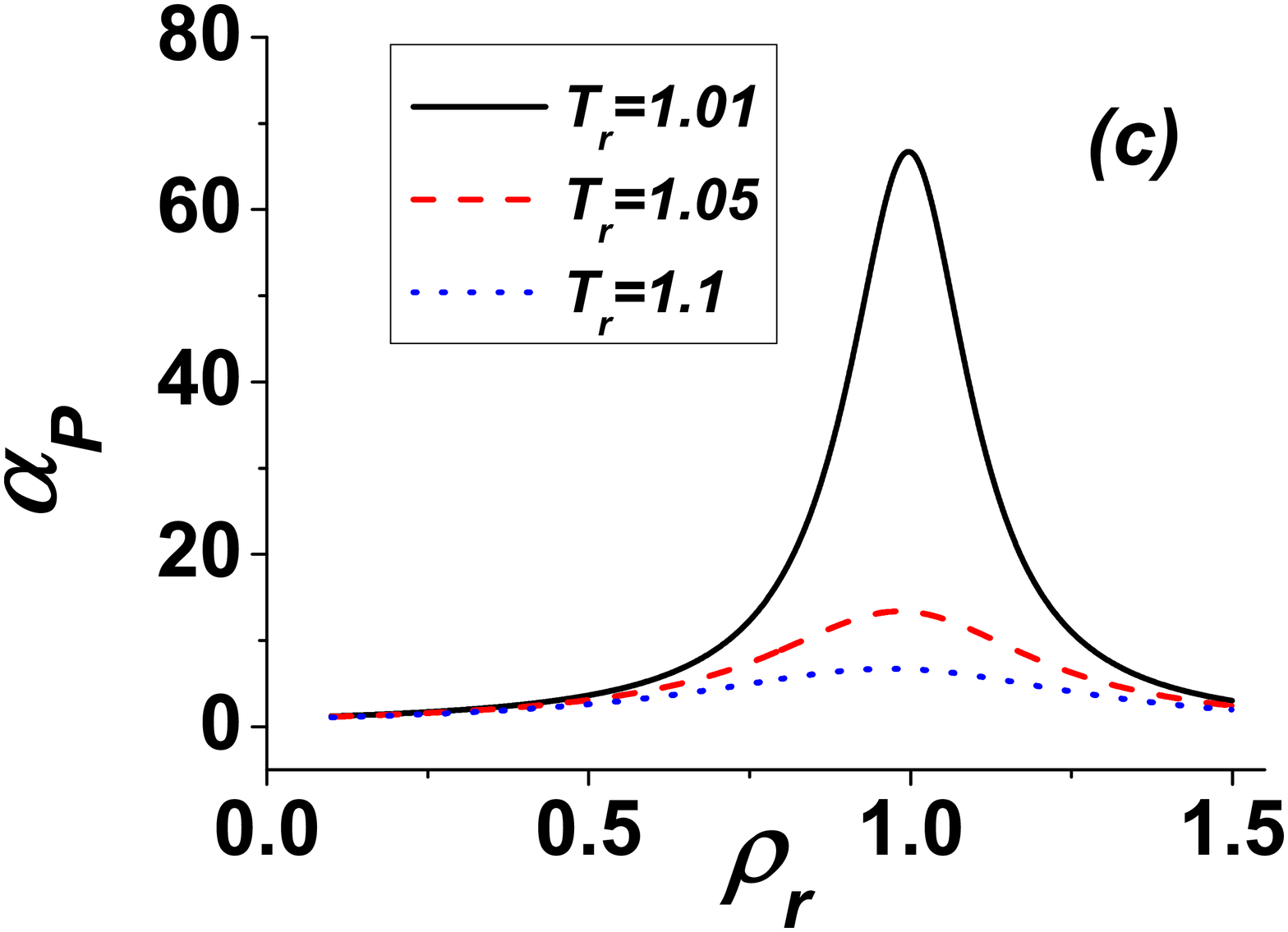}%
\includegraphics[width=4.5cm]{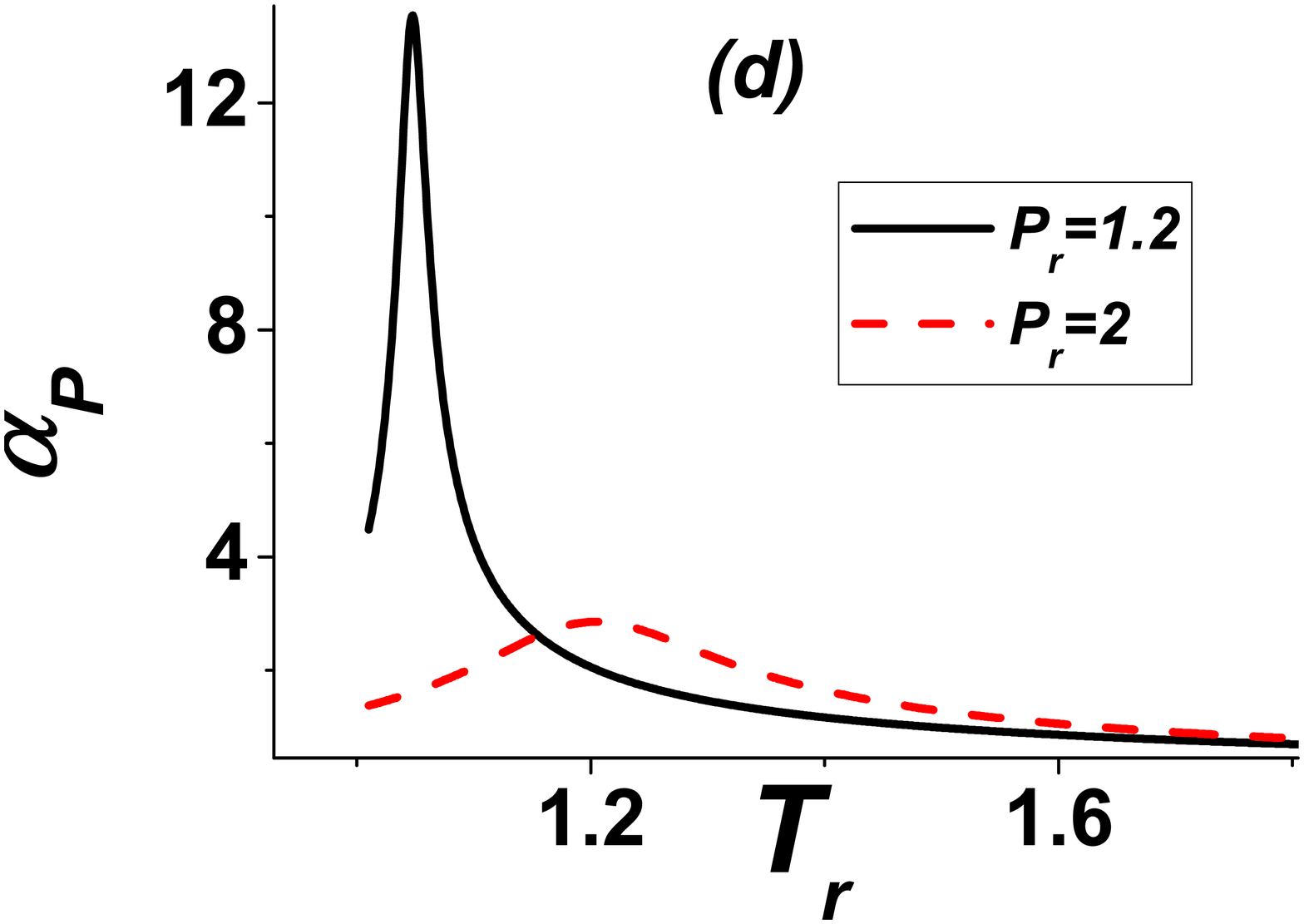}%

\includegraphics[width=4.5cm]{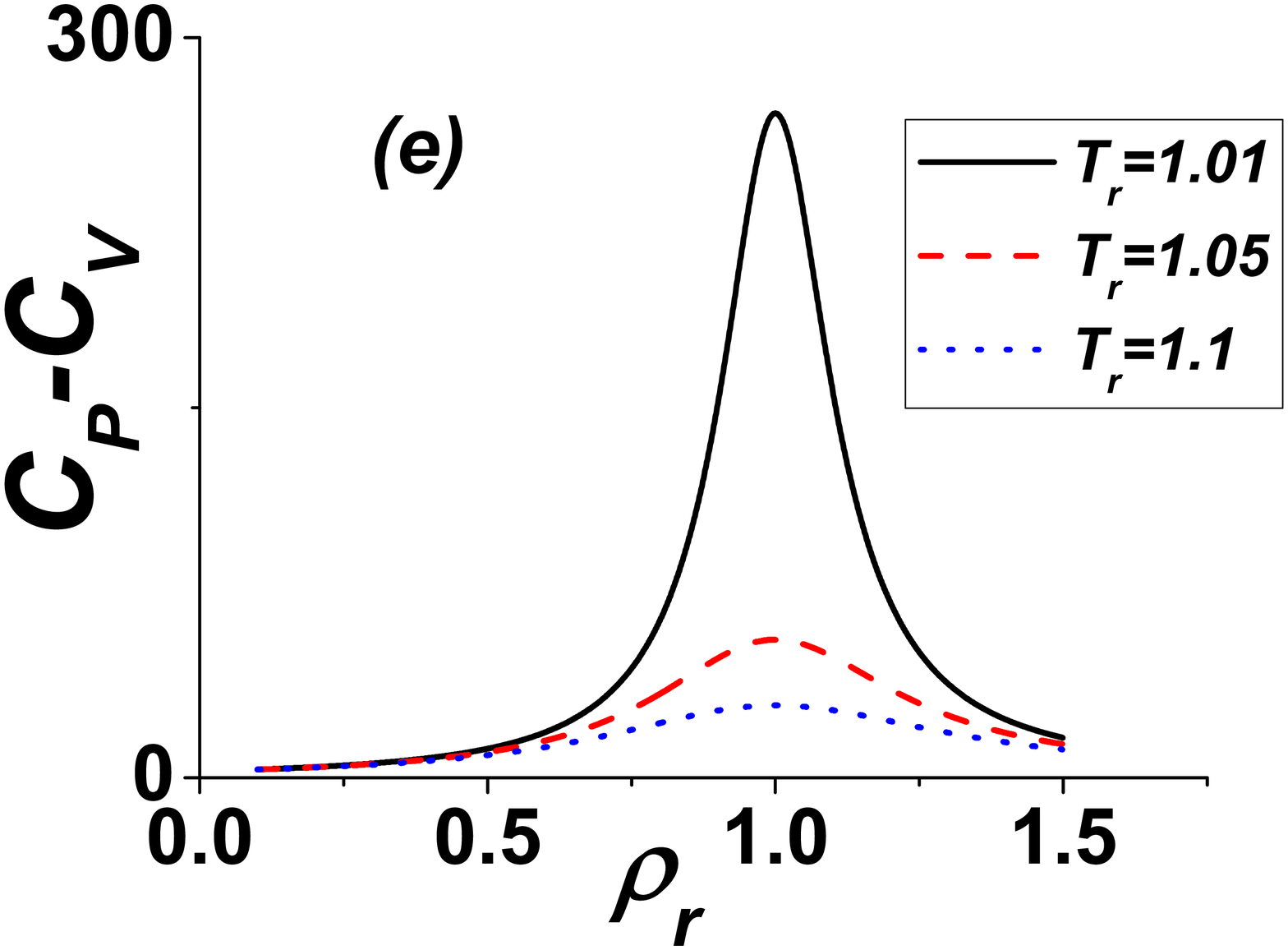}%
\includegraphics[width=4.5cm]{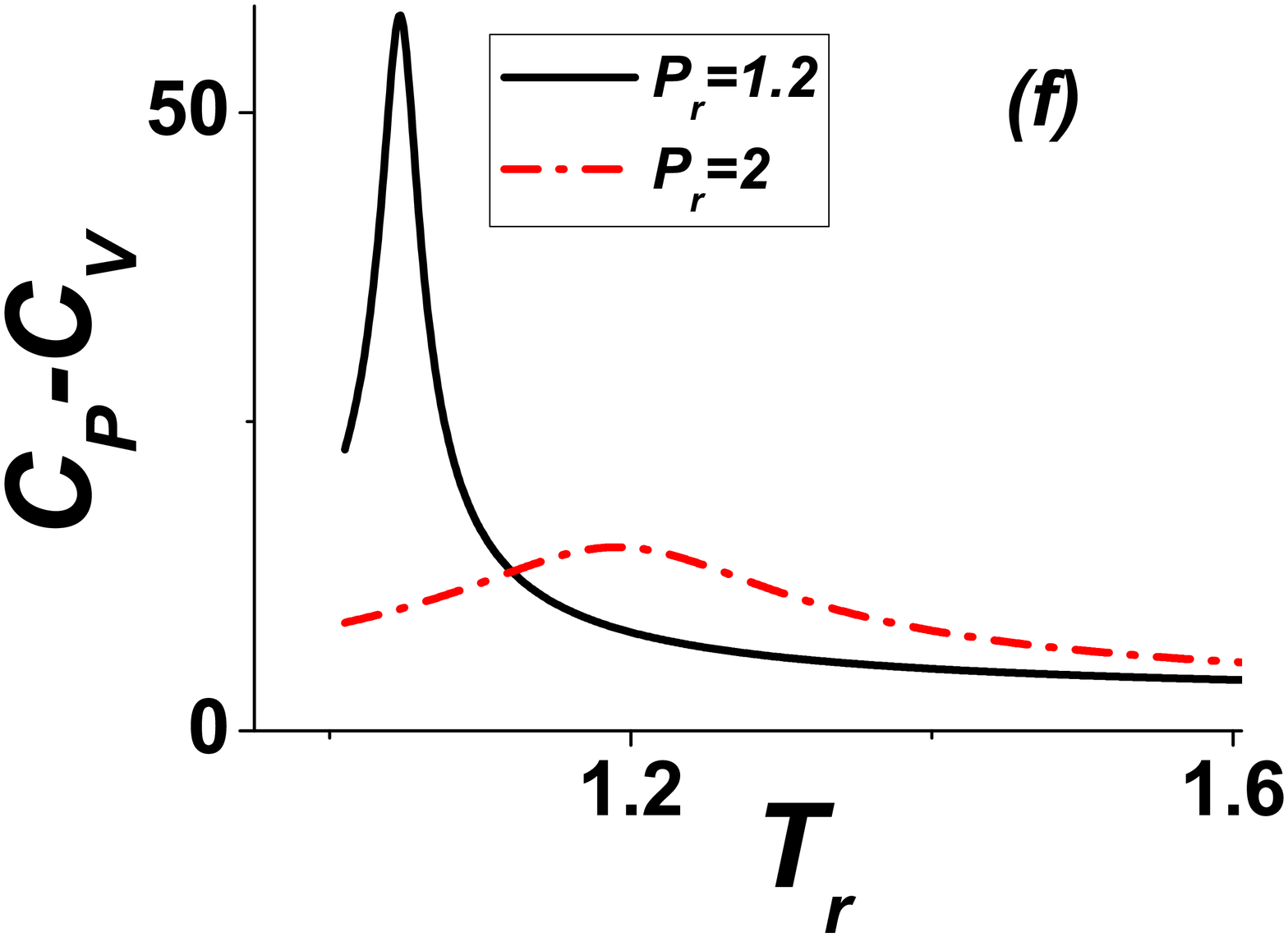}%

\caption{\label{fig:fig2} (Color online) Maxima of the thermal
expansion coefficient  $\alpha_P$, compressibility  $\beta_T$,
heat capacity $C_P-C_V$ in the isotherms ((a), (c), (e)) and
isobars ((b), (d), (f)).}
\end{figure}

Although all ridges are described by different equations, they are
close together near the critical point. For the estimate, the
lines of the extrema can be thought as coinciding, if the
temperature values on the lines at the same pressure differ by
less than 1\%. The value of 1\% roughly corresponds to the
experimental accuracy of a measurement of the respective values
and to the errors in the computer simulation data. For
van-der-Waals fluid the positions of all ridges for different
thermodynamic values merge into "single" Widom line in the $P,T$-
coordinates at $T<1.07 T_c$, $P<1.25 P_c$ (see the inset in
Fig.~\ref{fig:fig1}(b)). Although the line of the maxima of
density fluctuations and that of the maxima of thermal expansion
coefficient for a van-der-Waals fluid formally go till zero
pressures, and the line of the maxima of heat capacity goes to an
infinite temperature region, the amplitude of all extrema very
rapidly decays if we move away from the critical point (see
Fig.~\ref{fig:fig2}). It is of interest to notice the existence of
an additional minimum on the temperature dependence of
compressibility $\beta_T$ in the isobars (see the inset in
Fig.~\ref{fig:fig2}(b)).

As a criterion of actual disappearance of the extremum, one can
consider the ratio of a respective thermodynamic value in the
maximum or minimum to this value at densities being 10\% different
from the density in the extremum. If this ratio is below 1.01 (the
difference between the extremal value and the "background" value
is below 1\%), the "ridge" can be thought of as actually smeared.
When using the above criterion, the lines of all extrema, in fact,
end at rather moderate temperatures and pressures: $T_r ~\sim
1.59, P_r \sim  2.78, \rho_r \sim 0.83$ for $\alpha_P$; $T_r \sim
1.44, P_r \sim 2.13, \rho_r \sim  0.74$ for $\zeta_T$; $T_r \sim
1.73, P_r \sim  3.9, \rho_r = 1$ for $C_p$; $T_r \sim 1.15, P_r
\sim 1.35, \rho_r \sim 0.73$ for $\beta_T$; $T_r \sim 1.28, P_r
\sim 1.84, \rho_r \sim 0.83$ for $V_s$ (see inset in
Fig.~\ref{fig:fig1}(b)).

The lines of the maxima of most quantities correspond to a
decrease in density with increased temperature (see
Fig.~\ref{fig:fig1}); only the line of the maxima of the heat
capacity $Ñ_ð$ lies on the isochore.

Thus, a "thermodynamic" continuation of the gas-liquid phase
equilibrium line for a van-der-Waals fluid represents single Widom
line if the temperature moves only 7 \% away from the critical
point; if the temperature moves further away, it represents a
rapidly widening bunch of lines, which in fact ends at $T_r < 2$ è
$P_r < 5$.

\begin{figure}
\includegraphics[width=8cm]{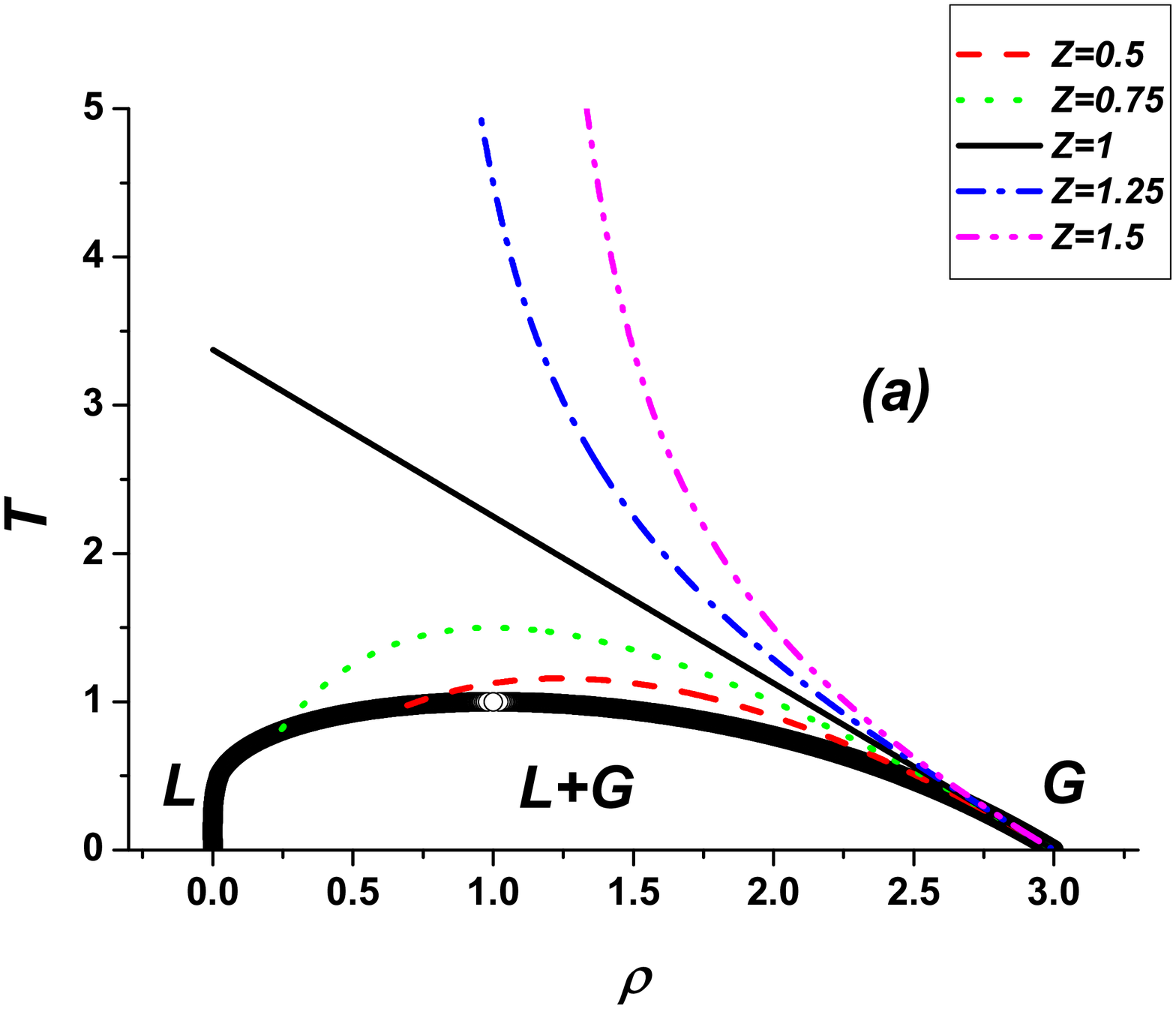}%

\includegraphics[width=8cm]{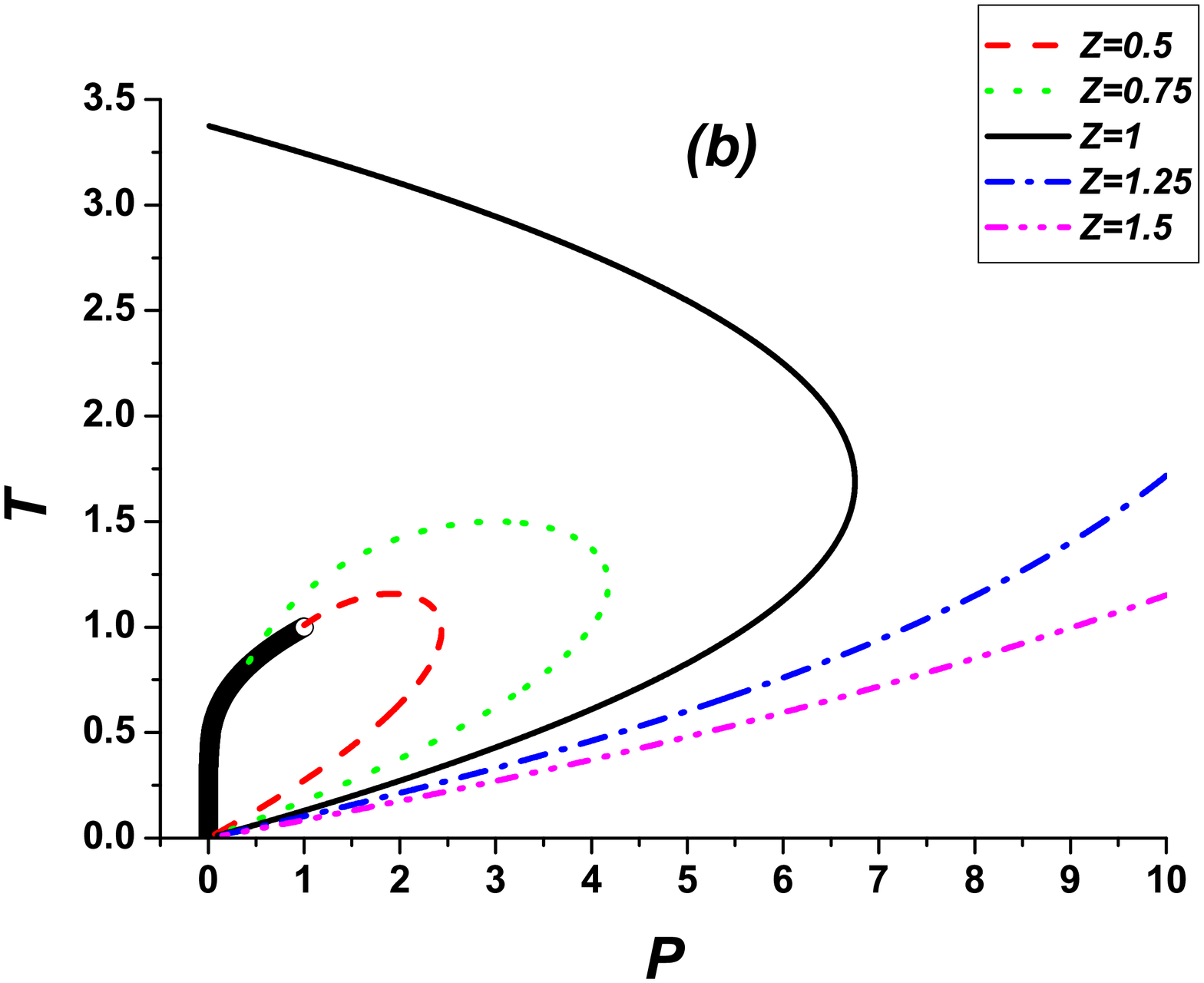}%

\caption{\label{fig:fig3} (Color online) The Batschinski lines in
the ($\rho-T$) (a) and ($P-T$) (b) planes. Thick lines correspond
to a liquid-gas transition.}
\end{figure}

Let us discuss now the Batschinski line, sometimes referred to as
the Zeno line. This line corresponds to the equation $PV/RT=1$.
A.I. Batschinski demonstrated that within the van-der-Waals model,
the above line will be straight in the coordinates $\rho,T$
\cite{[8]}. In most studies, the behavior of this line was only
analyzed in the $\rho,T$ coordinates for model and real systems
\cite{[9],[10]}. At the same time it is of interest to examine not
only the behavior of this line but also the behavior of other
lines determined by the condition $PV/RT=Z$ both in the $\rho,T$
and $T,P$-planes. Fig.~\ref{fig:fig3} presents the lines
satisfying the conditions $PV/RT=Z$ for a van-der-Waals fluid,
obtained from the equation:
\begin{equation}
T_r=\frac{3\rho_r(3-\rho_r)}{8-k(3-\rho_r)}. \label{bl}
\end{equation}
For the Batschinski line, we have the expression $T_r=\frac{9}{8}
(3-\rho_r)$. In this case, it should be taken into account that
the condition $PV/RT=Z$ in the reduced units has the form:
$\frac{P_r}{\rho_r T_r}=k$, where $k=8/3$ corresponds to the
condition $PV/RT=1$. The behavior of the lines if $PV/RT<1$ and
$PV/RT>1$ is quite different. The Batschinski line, determined by
the equation $PV/RT=1$, separates two regions of a fluid: a "soft"
low density fluid with the predominance of the attractive
potential as compared to an ideal gas, where $PV/RT<$1, and a more
"rigid" higher density fluid with the predominance of the
repulsive potential as compared to an ideal gas, where $PV/RT>1$
(see Fig.~\ref{fig:fig3}). Therefore, the line $PV=RT$ can be
called "separatrix". The Batschinski separatrix is the only line
of the given family which ends at zero density and pressure, where
it coincides with the ideal gas regime. The Batschinski separatrix
for the van-der-Waals fluid ends at $T_r=27/8$, i.e., the same
temperature at which the lines of the maxima of density
fluctuations and minima of the speed of sound end. This
coincidence is related to the fact that these lines correspond to
the zero of the second pressure derivative of the density. For an
ideal gas, this derivative is equal to zero at all temperatures,
so the line of the maxima of the value $\zeta_T$ at zero density
corresponds to the ideal gas equation $PV=RT$ as well.

Finally, let us analyze the behavior of a pseudo-Gruneisen
parameter for a van-der-Waals fluid. Using the equation (\ref{g})
we have:
\begin{equation}
\gamma= \frac{16}{3}\frac{\rho_r}{(3- \rho_r)}. \label{gvdw}
\end{equation}
For the majority of real substances, the Gruneisen parameter
varies in the range from 0.5 to 3, i.e., the van-der-Waals fluid
in the region of "normal" densities (2-2.5) for liquids has
anomalously high Gruneisen parameter values. For a system of
particles with a purely repulsive exponential interaction (soft
sphere system), the Gruneisen parameter value can easily be
deduced: $\gamma= (n+2)/6$, where $n$ - is the exponent of the
repulsive potential. Thus, $\rho_r=3$ limit obviously corresponds
to hard sphere system ($n=\infty$). At the critical point,
$\gamma=8/3\approx 2.67$, that is, coincides with the value for a
for soft sphere system with $n=14$, which is close to the exponent
in the repulsive part of Lennard-Jones potential ($n=12$). The
behavior of real rare gas substances, too, is well described by
the potential of soft spheres with $n=12-13$. Thus, the melting
curve for argon coincides, to a high accuracy, with that for a
soft sphere system with $n=12.7$ \cite{[17]}. Rare gas solids and
fluids also have typical values of the Gruneisen parameter
$\gamma\sim 2.3-2.6$ \cite{[12],[18]}. Thus, the van-der-Waals
fluid near the critical point has the Gruneisen parameter values
close to those for real rare gas fluids. In this regard, a success
in the description of the properties of rare gas fluids by
van-der-Waals equation in the vicinity of the critical point can
better be understood.

Summing up, one can conclude that in the framework of the
van-der-Waals equation it has been possible to obtain exact
analytical expressions for the "special" lines in the region of a
supercritical fluid and  for the pseudo-Gruneisen parameter. The
qualitative inferences made in the present study will obviously be
valid for real simple fluids as well. This is particularly true of
the conclusion about how far from the critical point we may
establish a single Widom line for different thermodynamic values,
and how far from the critical point the extrema of particular
physical quantities can still be followed. Of course, the behavior
of ridges for thermodynamic values may be different and depends on
corresponding equation of state \cite{[6],[7],[8n],[9n]}. However,
as for the liquid-gas transitions at enough high pressures ($P_r
>10$), it is dynamic, not thermodynamic, characteristics that
should be considered when separating a fluid into liquid-like and
gas-like regions.

\begin{acknowledgments}
The authors wish to thank S.M. Stishov, A.G. Lyapin and Yu.D.
Fomin for valuable discussions and referees for kind suggestions.
The work has been supported by the RFBR (No 11-02-00303, No
10-02-01407 No 11-02-00341), Russian Federal Programs
02.740.11.5160 and 02.740.11.0432), and by the Programs of the
Presidium of RAS.
\end{acknowledgments}

\end{document}